\newif\ifsupp
\supptrue 

\documentclass[superscriptaddress,twocolumn,10pt,prl,floatfix,nofootinbib]{revtex4-2}

\input{glyphtounicode}
\usepackage[T1]{fontenc}
\usepackage[utf8]{inputenc}
\usepackage[english]{babel}
\usepackage{amsmath}  
\usepackage{amsfonts} 
\usepackage{amssymb}
\usepackage{graphicx} 
\usepackage{hyperref}
\hypersetup{pdftex,bookmarks=true,bookmarksopen,bookmarksnumbered,
                colorlinks,
                linkcolor=blue,
                citecolor=blue}
\graphicspath{ {./} }
\usepackage{natbib}
\setcitestyle{super}
\usepackage{array}
\usepackage[table]{xcolor}
\usepackage{xcolor}
\usepackage{xr}
\usepackage{braket}
\usepackage{booktabs}
\usepackage{multirow}
\usepackage{dsfont} 
\usepackage{caption}

\usepackage{tabularx}
\usepackage{siunitx}

\usepackage{subcaption}
\usepackage{bigstrut}
\usepackage{svg}
\usepackage{upgreek}

\usepackage{newfloat}

\DeclareFloatingEnvironment[name={\textbf{Supplementary Figure}}]{suppfigure}

\newcommand{\aref}[1]{\hyperref[#1]{Appendix~\ref*{#1}}}
\DeclareCaptionLabelSeparator{line}{\hspace{2pt}\bf{|}\hspace{2pt}}

\begin{document}

\captionsetup[table]{name={\bf{Table}},labelsep=period,justification=raggedright,font=small,singlelinecheck=false}
\captionsetup[figure]{name={\bf{Figure}},labelsep=line,justification=raggedright,font=small,singlelinecheck=false}


\renewcommand{\equationautorefname}{Eq.}
\renewcommand{\figureautorefname}{Fig.}
\renewcommand*{\sectionautorefname}{Sec.}

\title{A 300\,mm foundry silicon spin qubit unit cell exceeding 99\,\% fidelity in all operations}

\author{\textbf{Paul Steinacker}$^+$}
\email{p.steinacker@unsw.edu.au}
\affiliation{School of Electrical Engineering and Telecommunications, University of New South Wales, Sydney, NSW 2052, Australia}
\author{\textbf{Nard Dumoulin Stuyck}$^+$}
\email{nard@diraq.com}
\author{Wee Han Lim}
\author{Tuomo Tanttu}
\author{MengKe Feng}
\affiliation{School of Electrical Engineering and Telecommunications, University of New South Wales, Sydney, NSW 2052, Australia}
\affiliation{Diraq, Sydney, NSW, Australia}
\author{Andreas Nickl}
\affiliation{School of Electrical Engineering and Telecommunications, University of New South Wales, Sydney, NSW 2052, Australia}
\author{Santiago Serrano}
\affiliation{School of Electrical Engineering and Telecommunications, University of New South Wales, Sydney, NSW 2052, Australia}
\affiliation{Diraq, Sydney, NSW, Australia}
\author{Marco Candido}
\affiliation{School of Electrical Engineering and Telecommunications, University of New South Wales, Sydney, NSW 2052, Australia}
\author{Jesus D. Cifuentes}
\author{Fay E. Hudson}
\author{Kok Wai Chan}
\affiliation{School of Electrical Engineering and Telecommunications, University of New South Wales, Sydney, NSW 2052, Australia}
\affiliation{Diraq, Sydney, NSW, Australia}
\author{Stefan Kubicek}
\author{Julien Jussot}
\author{Yann Canvel}
\author{Sofie Beyne}
\author{Yosuke Shimura}
\author{Roger Loo}
\author{Clement Godfrin}
\author{Bart Raes}
\author{Sylvain Baudot}
\author{Danny Wan}
\affiliation{Imec, Leuven, Belgium}
\author{Arne Laucht} 
\affiliation{School of Electrical Engineering and Telecommunications, University of New South Wales, Sydney, NSW 2052, Australia}
\affiliation{Diraq, Sydney, NSW, Australia}
\author{Chih Hwan Yang}
\affiliation{School of Electrical Engineering and Telecommunications, University of New South Wales, Sydney, NSW 2052, Australia}
\affiliation{Diraq, Sydney, NSW, Australia}
\author{Andre Saraiva}
\affiliation{Diraq, Sydney, NSW, Australia}
\author{Christopher C. Escott}
\affiliation{Diraq, Sydney, NSW, Australia}
\author{Kristiaan De Greve}
\affiliation{Imec, Leuven, Belgium}
\affiliation{Department of Electrical Engineering (ESAT), KU Leuven, Leuven, Belgium}
\author{Andrew S. Dzurak}
\email{andrew@diraq.com}
\affiliation{School of Electrical Engineering and Telecommunications, University of New South Wales, Sydney, NSW 2052, Australia}
\affiliation{Diraq, Sydney, NSW, Australia}

\date{\today}

\begin{abstract}
\boldmath
\textbf{Fabrication of quantum processors in advanced 300\,mm wafer-scale complementary metal–oxide–semiconductor (CMOS) foundries provides a unique scaling pathway towards commercially viable quantum computing with potentially millions of qubits on a single chip. Here, we show precise qubit operation of a silicon two-qubit device made in a 300\,mm semiconductor processing line. The key metrics including single- and two-qubit control fidelities exceed 99\,\% and state preparation and measurement fidelity exceeds 99.9\,\%, as evidenced by gate set tomography (GST). We report coherence and lifetimes up to $T_\mathrm{2}^{\mathrm{*}} = 30.4$ $\upmu$s, $T_\mathrm{2}^{\mathrm{Hahn}} = 803$ $\upmu$s, and $T_1 = 6.3$ s. Crucially, the dominant operational errors originate from residual nuclear spin carrying isotopes, solvable with further isotopic purification, rather than charge noise arising from the dielectric environment. Our results answer the longstanding question whether the favourable properties including high-fidelity operation and long coherence times can be preserved when transitioning from a tailored academic to an industrial semiconductor fabrication technology.}
\unboldmath
\end{abstract}
\maketitle


\def\thefootnote{+}\footnotetext{These authors contributed equally to this work.}\def\thefootnote{\arabic{footnote}}


The early academic successes of spin qubit technology in silicon devices~\cite{Veldhorst2015,Huang2019, noiri_fast_2022,madzik_precision_2022,xue_quantum_2022,mills_two-qubit_2022, weinstein_universal_2023, Yang2020,huang_high-fidelity_2024,tanttu_assessment_2024} created enthusiasm about the fabrication of qubits in industrial \SI{300}{\milli\meter} manufacturing lines~\cite{dumoulin_stuyck_cmos_2024,zwerver_qubits_2022,maurand_cmos_2016-1,camenzind_hole_2022}. The potential advantages include leveraging the advanced fabrication processes, integrating qubits with modern electronics capabilities, and benefiting from the installed capabilities for mass manufacturing at economical costs. 


Recent progress on harmonising spins in silicon quantum dots with industrial-scale fabrication raised awareness about the materials challenges that need to be faced~\cite{saraiva_materials_2022,elsayed_low_2024,neyens_probing_2024}. High-fidelity single-qubit operations can be performed with less impact from charge noise because the electron spin resonance only couples to the noisy electric fields through its weak spin-orbit effects~\cite{dumoulin_stuyck_demonstration_2024, jock_silicon_2018, tanttu_controlling_2019, burkard_semiconductor_2023, cifuentes_impact_2024}. In contrast, other operations including qubit initialisation, readout and exchange-based two-qubit gates are more sensitive to charge noise present in traditional metal-oxide-semiconductor (MOS) gate stacks~\cite{burkard_semiconductor_2023,elsayed_low_2024}. 

This motivated the search for alternative approaches including moving spin states away from the interface to quantum wells~\cite{lawrie_quantum_2020,scappucci_germanium_2021}. From a scaling perspective, however, it is desirable to reconcile traditional CMOS processes with qubit fabrication to leverage on the mature CMOS industry. Moreover, the formation of quantum dots against thin oxides has several benefits, such as a large electric field that leads to strong valley splitting~\cite{yang_spin-valley_2013,Cifuentes2024-du,saraiva_materials_2022} and efficient electrostatic coupling between gate voltage bias and the channel, allowing for well-isolated states, efficient exchange coupling control~\cite{Cifuentes2024-du}, and fast gate-based readout~\cite{gonzalez-zalba_gate-sensing_2016}. \\

In this work, we study qubits fabricated using a long-established CMOS geometry, namely planar metal-oxide-insulator with polysilicon gates~\cite{li_flexible_2020}, and demonstrate all the basic qubit operations, including one- and two-qubit gates, initialisation and readout, with error rates approaching the fault-tolerance threshold for the widely studied surface error correction code~\cite{fowler_surface_2012,shaw2022quantum}. These error rates are studied with state-of-the-art gate set tomography (GST) tools~\cite{blume-kohout_demonstration_2017,nielsen_probing_2020} to pinpoint effects such as crosstalk and the breakdown between stochastic and coherent errors. 

\section{Device fabrication and operation}

\begin{figure*}
   \includegraphics[width =\textwidth]{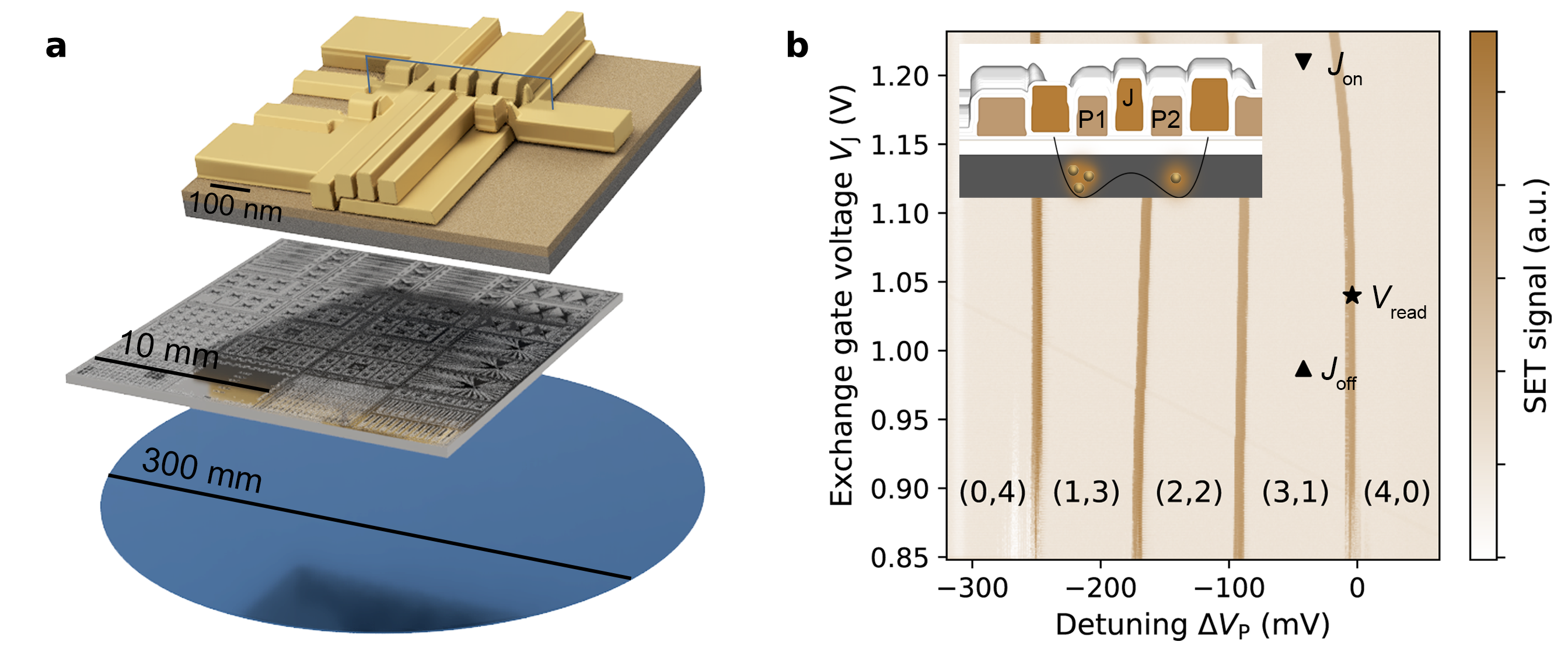}
    \caption{\textbf{Two-qubit device and operation points.} 
    \textbf{a}, Schematic of a Diraq two-qubit device fabricated on a \SI{300}{\milli\meter} wafer, showing the full wafer, single die, and single device level.
    \textbf{b}, Charge stability diagram as a function of plunger (P1, P2) voltage detuning $\Delta V_\mathrm{P} = \Delta V_\mathrm{P1} = -\Delta V_\mathrm{P2}$ and exchange gate (J) voltage $V_\mathrm{J}$, showing four isolated electrons in the double-dot system. Voltage points for single-qubit operation ($J_{\rm off}$), two-qubit operation ($J_{\rm on}$) and readout are labelled as triangle (\scalebox{1.0}{$\blacktriangle$}), triangle down (\scalebox{1.0}{$\blacktriangledown$}), and star (\scalebox{0.80}{$\bigstar$}), respectively. \textbf{Inset:} Cross section of the double quantum dot device indicating electron numbers under the plunger gates (P1,P2), exchange gate (J), and double potential well.  
}
    \label{fig:fig1}
\end{figure*}

The qubit device is designed by Diraq and fabricated at imec in a $\SI{300}{\milli\meter}$ spin qubit process flow optimised for low charge-noise and high-uniformity~\cite{dumoulin_stuyck_integrated_2020, li_flexible_2020, dumoulin_stuyck_uniform_2021, elsayed_low_2024}. The process flow is optimised for qubit specific integration with gate pitch smaller than $\SI{100}{\nano\meter}$. Fast fabrication cycles and design flexibility are ensured by optimally combining optical lithography and electron beam lithography (EBL). The fabrication starts with epitaxial growth of an isotopically enriched layer of silicon with a residual concentration of \SI{400}{ppm} $^{29}$Si. A high-quality thermally grown oxide forms the Si/SiO$_2$ interface at which the charge of the electron spin qubit is accumulated. Next, a triple-layer overlapping polysilicon gate stack is formed using EBL and etch processes, with each gate separated by a thin interstitial high-temperature oxide deposition~\cite{elsayed_low_2024}. Finally, an aluminium stripline to manipulate single spin states using electron spin resonance (ESR) is patterned in proximity to the device as an intermediate control solution at the few-qubit level~\cite{vahapoglu_single-electron_2021,vahapoglu_coherent_2022}.

The design consists of a double quantum dot and a nearby single-electron transistor (SET) for spin readout~\cite{Veldhorst2015} (Fig.~\ref{fig:fig1}a). We operate the device in a $^3$He/$^4$He dilution refrigerator with a base temperature of 10 mK in isolation mode with four electrons in the double dot formed under the plunger gate electrodes (P1, P2)~\cite{Yang2020}. An interstitial exchange gate (J) voltage-controls the tunnel coupling between the dots~\cite{Veldhorst2015, Chittock-Wood2024-hp}. The electrons are loaded from a two-dimensional electron gas formed under the reservoir gate which overlaps with an n$^{++}$ implanted ohmic region. Pauli Spin Blockade (PSB) at the (3,1)-(4,0) charge transition is used for spin parity readout and the signal is measured using the SET in dc mode~\cite{seedhouse_pauli_2021} (Fig.~\ref{fig:fig1}b). SET charge noise measurements reveal a $1/f^\alpha$ noise spectrum with $\alpha \approx 0.2$ and an amplitude of approximately $\SI{0.8}{\micro\electronvolt\per \sqrt{\hertz}}$ at $\SI{1}{\hertz}$, in line with previous low-charge noise results from similarly fabricated devices~\cite{elsayed_low_2024,dumoulin_stuyck_demonstration_2024} (Extended Data Fig.~\ref{fig:sup_fig3}).

\begin{figure}
    \includesvg[width = \columnwidth]{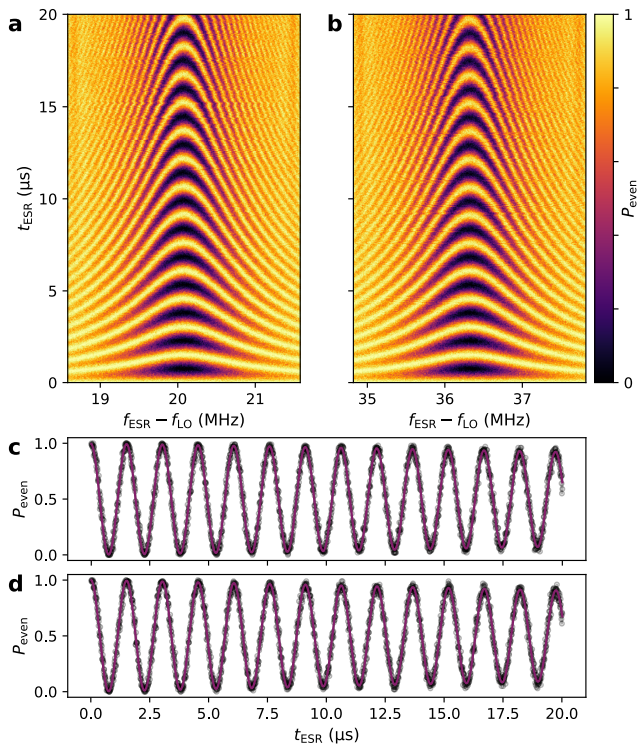}
    \caption{\textbf{Single qubit operation.} 
    $\textbf{a, b,}$ Rabi chevron for qubit 1 and 2, respectively. $\textbf{c, d,}$ Coherent Rabi oscillations at $f_\text{Larmor}$ for qubit 1 and 2, respectively. Real-time feedback is implemented to counteract Larmor frequency deviation~\cite{dumoulin_stuyck_silicon_2024}. The local oscillator frequency is set to $f_{\mathrm{LO}} = \SI{18.610}{\giga \hertz}$. The applied microwave power to Qubit 1 is approximately \SI{7}{\percent} larger resulting in matching Rabi frequencies of $f_\mathrm{Rabi} = \SI{658.6 \pm 0.3}{\kilo \hertz}$.
}
    \label{fig:fig2}
\end{figure}

A static magnetic field $B_0 = \SI{0.666}{\tesla}$ splits the spins by the Zeeman energy $E_\text{Z} \approx \SI{77}{\micro\electronvolt}$ to form the single qubit states which are manipulated with on-resonance microwave signals applied to the ESR antenna. Single qubit $X_{\pi/2}$ gates are formed with timed microwave pulses on resonance with the qubit Larmor frequencies $f_\text{Larmor}$ (Fig.~\ref{fig:fig2}), while $Z_{\pi/2}$ gates are implemented as virtual gates by frame rotations of the reference frame~\cite{vandersypen_nmr_2005}. Table~\ref{tab:table1} summarises the single qubit characterisation metrics for spin lifetime, Ramsey, and Hahn experiments. 

\begin{table}
    \caption{\textbf{Single qubit metrics.} Spin lifetime and coherence times from Ramsey and Hahn experiments (Extended Data Fig.~\ref{fig:sup_fig1}).
    Errors are based on the \SI{95}{\percent} confidence level.
}
    \centering
    \begin{tabular}{l l l}
        Metric & Qubit 1 & Qubit 2 \\
        \hline
        $T_1$ ($\SI{}{\second}$) & 2.4(2) & 6.3(6)  \\
        $T_\mathrm{2}^{\mathrm{*}}$ ($\SI{}{\micro\second}$) & 30.4(8)   & 29.1(6)  \\
        $T_\mathrm{2}^{\mathrm{Hahn}}$ ($\SI{}{\micro\second}$) & 445(6) & 803(6) \\
        \hline
    \end{tabular}\label{tab:table1}
\end{table}

\begin{figure*}
    \includesvg[width = \textwidth]{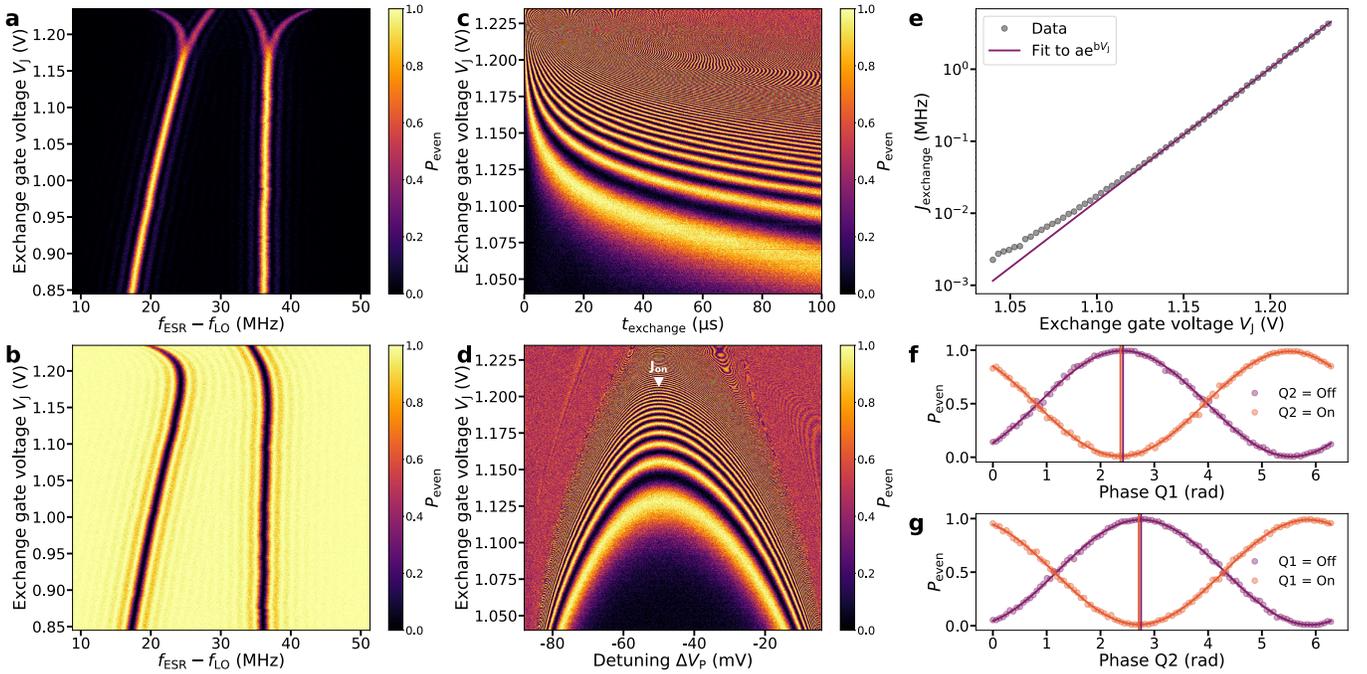}
    \caption{\textbf{Two qubit operation.} 
    \textbf{a, b}, Probability of detecting an even spin state, $P_\mathrm{even}$, after a microwave burst of fixed power and duration at different J gate voltages $V_\mathrm{J}$ when preparing a mixed odd state $\frac{1}{\sqrt{2}}(\ket{\downarrow\uparrow}+\ket{\uparrow\downarrow})$ (\textbf{a}), and a pure state $\ket{\downarrow\downarrow}$ (\textbf{b}), respectively. The power and duration of the microwave burst are roughly calibrated to a single-qubit $\pi$-rotation. The local oscillator frequency is set to $f_{\mathrm{LO}} = \SI{18.610}{\giga \hertz}$. The following experiments are conducted with $\ket{\downarrow\downarrow}$ initialisation, unless otherwise specified.
    \textbf{c} Controlled phase (CZ) oscillation as a function of exchange time $t_\mathrm{exchange}$ and exchange gate voltage $V_\mathrm{J}$. We apply dynamical decoupling by a consecutive $\pi$-rotation on both qubits to filter out other phase accumulating effects such as AC Stark shift.
    \textbf{d} Decoupled exchange oscillation fingerprint for fixed exchange time $t_\mathrm{exchange} = \SI{10}{\micro \second}$ as a function of plunger voltage detuning $\Delta V_\mathrm{P}$ and exchange gate voltage $V_\mathrm{J}$. Triangle indicates the exchange gate voltage used for the CZ gate.
    \textbf{e}, Exchange strength as a function of exchange gate voltage extracted from fitting exchange oscillations in \textbf{c}. The controllability is fitted assuming $J_\mathrm{Exchange}\propto a\exp{bV_\mathrm{J}}$ resulting in $b = \SI{18.4}{dec\per\volt}$. 
    \textbf{f, g}, Calibration of the CZ single qubit phase correction by preparing the target spin in superposition, applying a CPHASE followed by a virtual phase rotation for qubit 1 and 2, respectively~\cite{xue_quantum_2022,huang_high-fidelity_2024}. Vertical lines indicate the phase values where the spin state of Q1 (Q2) is flipped if Q2 (Q1) is in the on/spin up state. 
    Readout probability is unscaled in all data. 
}
    \label{fig:main_fig3}
\end{figure*}

Figure~\ref{fig:main_fig3} details the two-qubit control implementation. A voltage pulse, typically of length $t_{\text{CZ}}$ between 100 and \SI{500}{\nano \second}, applied to the J gate controls the exchange interaction strength (Fig.~\ref{fig:main_fig3}a-c). We perform the two-qubit gate at the symmetric charge operation point to reduce the sensitivity to the detuning noise~\cite{Reed2016} (Fig.~\ref{fig:main_fig3}d). We extract from the fitted exchange frequencies in Fig.~\ref{fig:main_fig3}e an exchange controllability of 18.4 dec/V$_\text{J}$. The combination of a pulsed exchange interaction together with single qubit phase rotations allows the implementation of a controlled-$Z$ (CZ) two-qubit gate ~\cite{Veldhorst2015, Yang2020} (Fig.~\ref{fig:main_fig3}f,g). Realtime feedback tracks and corrects the SET voltage operation point, and the single qubit Larmor and Rabi frequencies~\cite{dumoulin_stuyck_silicon_2024}. A heralded initialisation protocol checks if the qubit system was successfully initialised in the $\ket{\downarrow \downarrow}$ state and repeats the initialisation if unsuccessful~\cite{huang_high-fidelity_2024, steinacker_violating_2024}. 

\section{Two-qubit benchmarking}

\begin{figure*}
    \includesvg[width = \textwidth]{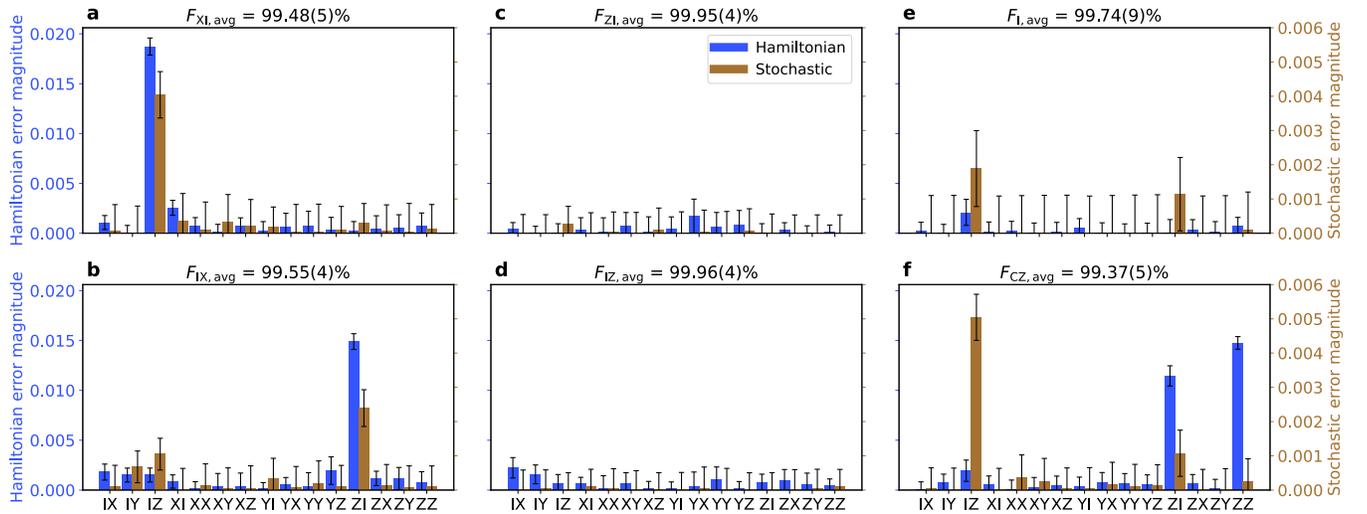}
    \caption{\textbf{Two-qubit benchmarking using GST.} 
    \textbf{a-f}, Magnitude of Hamiltonian and stochastic error generators of the full gate set  $\mathrm{XI}$ (\textbf{a}), $\mathrm{IX}$ (\textbf{b}), $\mathrm{ZI}$ (\textbf{c}), $\mathrm{IZ}$ (\textbf{d}), $\mathrm{I}$ (\textbf{e}), and $\mathrm{CZ}$ (\textbf{f}) from GST. Hamiltonian errors contribute to the infidelity in second order, while stochastic errors contribute in first order. The average gate fidelity is given above each plot. The on-target X gate fidelity $F_\mathrm{X,Qi}$ is calculated from the relevant error components. 
    Error bars represent the \SI{95}{\percent} confidence level. 
}
    \label{fig:main_fig4}
\end{figure*}

To precisely capture the errors in qubit operation we designed a GST experiment comprising the full gateset \{I, XI, IX, ZI, IZ, CZ\}, from which we construct a list of preparation and measurement fiducials. Combining these fiducials with germs up to length 16, we obtain a list of about 12000 unique sequences. We chose GST as a benchmarking method since it yields the error taxonomy and is generally a higher bar for the fidelities than the more commonly used interleaved randomized benchmarking which often overestimates the fidelities~\cite{Nielsen2021gatesettomography, tanttu_assessment_2024}.

The idle gate is implemented as a wait of length $t_\text{CZ}$. The two-qubit XI and IX gate, as well as the virtual ZI and IZ gate are single-qubit $\pi/2$ rotations around the $\hat{x}$- and $\hat{z}$-axes for qubit 1 and 2, respectively, while the other qubit is idling. The two-qubit entangling CZ gate is implemented by turning on exchange at $V_\mathrm{J} = \SI{1.21}{\volt}$ for $t_\text{CZ} = \SI{212}{\nano \second}$. 

Figures~\ref{fig:main_fig4}a-f break down the error magnitude components of the full gate set. All operations are above the $\SI{99}{\percent}$ threshold with gate fidelities of $F_\mathrm{XI,avg} = \SI{99.45\pm 0.05}{\percent}$, $F_\mathrm{IX,avg} = \SI{99.55\pm 0.04}{\percent}$, $F_\mathrm{ZI,avg} = \SI{99.95\pm 0.04}{\percent}$, $F_\mathrm{IZ,avg} = \SI{99.96\pm 0.04}{\percent}$, $F_\mathrm{I,avg} = \SI{99.74\pm 0.09}{\percent}$, and $F_\mathrm{CZ,avg} = \SI{99.37\pm 0.05}{\percent}$. The combined state preparation and measurement fidelity is even above the $\SI{99.9}{\percent}$ threshold with $F_\mathrm{SPAM} = \SI{99.95\pm 0.08}{\percent}$ (Extended Data Fig.~\ref{fig:sup_fig4}). Single-qubit on-target gate fidelities of $F_\mathrm{X,Q1} = \SI{99.97 \pm 0.15}{\percent}$ and $F_\mathrm{X,Q2} = \SI{99.81 \pm 0.11}{\percent}$ are comparable to previously reported values from devices made using the same fabrication processes~\cite{dumoulin_stuyck_demonstration_2024}.

A number of error channel contributions and trends stand out. Firstly, the stochastic error magnitude is highest for the IZ error channel across the XI, IX, I and CZ gates (Fig.~\ref{fig:main_fig4}). A possible explanation is the coupling between a nuclear spin state and the electron spin state of qubit 2~\cite{Muhonen2014}. The Larmor frequency variation, tracked during the GST experiment using real-time feedback~\cite{dumoulin_stuyck_silicon_2024, steinacker_violating_2024}, shows a higher standard deviation for the qubit 2 frequency variation compared to qubit 1 (Extended Data Fig.~\ref{fig:sup_fig2}). Secondly, charge noise would primarily manifest in the CZ gate stochastic error magnitude ZZ error channel~\cite{tanttu_assessment_2024}. The data show, however, that the two main contributions for the CZ gate are ZI and IZ (the latter being the highest). The nuclear spin states of the residual $^{29}$Si atoms which couple strongly to qubit~2 are the likely cause for this effect~\cite{Huang2019,hensen_silicon_2020,zhao_single-spin_2020}.

The observation that the operational fidelity above degrades mostly due to noise in qubit~2 which seems contradictory, since $T_\mathrm{2,Q2}^{\mathrm{Hahn}}>T_\mathrm{2,Q1}^{\mathrm{Hahn}}$, $T_\mathrm{2,Q2}^{\mathrm{*}}\approx T_\mathrm{2,Q1}^{\mathrm{*}}$, and $T_\mathrm{1,Q2}>T_\mathrm{1,Q1}$. This is likely since, for the operation, the dominant errors in GST come from the qubit Larmor jumps which are not fully compensated by the Larmor feedback (Extended Data Fig.~\ref{fig:sup_fig2}). These slow time scale (likely $^{29}$Si) jumps do not contribute to the $T_\mathrm{2}^{\mathrm{Hahn}}$ or to the spin-phonon coupling that limit $T_\mathrm{1}$. We believe that the stronger spin-orbit interaction in qubit~1 compared to qubit~2 seen in Fig.~\ref{fig:main_fig3}ab combined with the charge noise limits $T_\mathrm{2,Q1}^{\mathrm{Hahn}}$ and $T_\mathrm{1,Q1}$. This is though not at the level that would be dominant in our GST error assessment.

The ZI and IZ gates are implemented virtually and in principle should not have any error. We attribute their infidelities, $\SI{0.05}{\percent}$ ($\SI{0.04}{\percent}$) for ZI (IZ), to a small time delay in the FPGA-based control hardware while updating the reference frame resulting in a stochastic error contribution on their ZI and IZ error channels.

\section{Discussion}

One of the most widely identified main limiting factors to achieving high control fidelity and coherence with spin qubits is charge noise~\cite{tanttu_assessment_2024, xue_quantum_2022, mills_two-qubit_2022,noiri_fast_2022,Struck2020, Reed2016,Yoneda2018}.
The results in this work are strongly encouraging from a noise analysis perspective, as several features in our data point toward the residual nuclear spin in the Si substrate, and not the charge noise, being the fidelity limiting factor. We plan to further investigate the fidelity limiting factors in future experiments. A clear pathway to reduce nuclear spin noise is further isotopical enrichment from the \SI{400}{ppm} $^{29}$Si used in this work to levels at or below \SI{50}{ppm} as already demonstrated in academic prototype devices~\cite{tyryshkin_electron_2012,zhao_single-spin_2020}.

We demonstrate that with engineering practices in an industrial setting the systematic search for target performance metrics can lead to significant improvements to the state-of-the-art in spin qubit performance. It is crucial to understand the connection between qubit performance and measurements that can be done inline with the fabrication process, such as electrical noise and Hall bar transport~\cite{elsayed_low_2024, camenzind_high_2021}. These investigations are in their infancy and will require an increasingly accurate model of the quantum behaviour of electrons in semiconductor devices.


Improvements on the qubit front will also be required. Firstly, these initial demonstrations are not yet at a level of fidelity that is conducive to scalable quantum computing with reasonable overheads~\cite{fowler_surface_2012}. With the new challenges of large-scale integration, qubit control must be revised and a good target is \SI{99.9}{\percent} fidelities across all operations, which  SPAM already reaches here.

At this stage, a statistical analysis of these qubits is not yet available. The consistent calibration of qubits at the level demonstrated here remains an intensive manual process with a number of scientific insights gained along the way. Automation of this process needs to mature in order to consistently achieve these results in a mass-characterisation campaign. \\

These results can also be leveraged to design quantum error correction strategies to cater to the specific demands of spin qubits. However, further studies will be needed to characterise qubit operations in a more scalable setting, such as under a single global microwave field and under elevated temperatures, which will inevitably be imposed by the heat dissipated by the co-integrated control electronics on the CMOS chip~\cite{vahapoglu_coherent_2022,bartee_spin_2024}. The paths found here in this study for generating high quality qubits in CMOS \SI{300}{\milli\meter} foundries should encourage further industrial research into far more complex technologies that are at the disposal of foundries.

\bibliography{mybib}

\begin{thebibliography}{10}
\expandafter\ifx\csname url\endcsname\relax
  \def\url#1{\texttt{#1}}\fi
\expandafter\ifx\csname urlprefix\endcsname\relax\def\urlprefix{URL }\fi
\providecommand{\bibinfo}[2]{#2}
\providecommand{\eprint}[2][]{\url{#2}}

\bibitem{Veldhorst2015}
\bibinfo{author}{Veldhorst, M.} \emph{et~al.}
\newblock \bibinfo{title}{A two-qubit logic gate in silicon}.
\newblock \emph{\bibinfo{journal}{Nature}} \textbf{\bibinfo{volume}{526}}, \bibinfo{pages}{410--414} (\bibinfo{year}{2015}).

\bibitem{Huang2019}
\bibinfo{author}{Huang, W.} \emph{et~al.}
\newblock \bibinfo{title}{Fidelity benchmarks for two-qubit gates in silicon}.
\newblock \emph{\bibinfo{journal}{Nature}} \textbf{\bibinfo{volume}{569}}, \bibinfo{pages}{532--536} (\bibinfo{year}{2019}).
\newblock \urlprefix\url{https://doi.org/10.1038/s41586-019-1197-0}.

\bibitem{noiri_fast_2022}
\bibinfo{author}{Noiri, A.} \emph{et~al.}
\newblock \bibinfo{title}{Fast universal quantum gate above the fault-tolerance threshold in silicon}.
\newblock \emph{\bibinfo{journal}{Nature}} \textbf{\bibinfo{volume}{601}}, \bibinfo{pages}{338} (\bibinfo{year}{2022}).
\newblock \urlprefix\url{https://doi.org/10.1038/s41586-021-04182-y}.

\bibitem{madzik_precision_2022}
\bibinfo{author}{Mądzik, M.~T.} \emph{et~al.}
\newblock \bibinfo{title}{Precision tomography of a three-qubit donor quantum processor in silicon}.
\newblock \emph{\bibinfo{journal}{Nature}} \textbf{\bibinfo{volume}{601}}, \bibinfo{pages}{348} (\bibinfo{year}{2022}).
\newblock \urlprefix\url{https://doi.org/10.1038/s41586-021-04292-7}.

\bibitem{xue_quantum_2022}
\bibinfo{author}{Xue, X.} \emph{et~al.}
\newblock \bibinfo{title}{Quantum logic with spin qubits crossing the surface code threshold}.
\newblock \emph{\bibinfo{journal}{Nature}} \textbf{\bibinfo{volume}{601}}, \bibinfo{pages}{343} (\bibinfo{year}{2022}).
\newblock \urlprefix\url{https://doi.org/10.1038/s41586-021-04273-w}.

\bibitem{mills_two-qubit_2022}
\bibinfo{author}{Mills, A.~R.} \emph{et~al.}
\newblock \bibinfo{title}{Two-qubit silicon quantum processor with operation fidelity exceeding 99\%}.
\newblock \emph{\bibinfo{journal}{Science Advances}} \textbf{\bibinfo{volume}{8}}, \bibinfo{pages}{5130} (\bibinfo{year}{2022}).
\newblock \urlprefix\url{https://www.science.org/doi/full/10.1126/sciadv.abn5130}.

\bibitem{weinstein_universal_2023}
\bibinfo{author}{Weinstein, A.~J.} \emph{et~al.}
\newblock \bibinfo{title}{Universal logic with encoded spin qubits in silicon}.
\newblock \emph{\bibinfo{journal}{Nature}} \textbf{\bibinfo{volume}{615}}, \bibinfo{pages}{817--822} (\bibinfo{year}{2023}).
\newblock \urlprefix\url{https://www.nature.com/articles/s41586-023-05777-3}.

\bibitem{Yang2020}
\bibinfo{author}{Yang, C.~H.} \emph{et~al.}
\newblock \bibinfo{title}{Operation of a silicon quantum processor unit cell above one kelvin}.
\newblock \emph{\bibinfo{journal}{Nature}} \textbf{\bibinfo{volume}{580}}, \bibinfo{pages}{350--354} (\bibinfo{year}{2020}).
\newblock \urlprefix\url{http://www.nature.com/articles/s41586-020-2171-6}.

\bibitem{huang_high-fidelity_2024}
\bibinfo{author}{Huang, J.~Y.} \emph{et~al.}
\newblock \bibinfo{title}{High-fidelity spin qubit operation and algorithmic initialization above 1 {K}}.
\newblock \emph{\bibinfo{journal}{Nature}} \textbf{\bibinfo{volume}{627}}, \bibinfo{pages}{772--777} (\bibinfo{year}{2024}).
\newblock \urlprefix\url{https://www.nature.com/articles/s41586-024-07160-2}.

\bibitem{tanttu_assessment_2024}
\bibinfo{author}{Tanttu, T.} \emph{et~al.}
\newblock \bibinfo{title}{Assessment of the errors of high-fidelity two-qubit gates in silicon quantum dots}.
\newblock \emph{\bibinfo{journal}{Nature Physics}} \bibinfo{pages}{1--6} (\bibinfo{year}{2024}).
\newblock \urlprefix\url{https://www.nature.com/articles/s41567-024-02614-w}.

\bibitem{dumoulin_stuyck_cmos_2024}
\bibinfo{author}{Dumoulin~Stuyck, N.} \emph{et~al.}
\newblock \bibinfo{title}{Cmos compatibility of semiconductor spin qubits} (\bibinfo{year}{2024}).
\newblock \urlprefix\url{https://arxiv.org/abs/2409.03993}.
\newblock \eprint{arXiv:2409.03993}.

\bibitem{zwerver_qubits_2022}
\bibinfo{author}{Zwerver, A.~M.} \emph{et~al.}
\newblock \bibinfo{title}{Qubits made by advanced semiconductor manufacturing}.
\newblock \emph{\bibinfo{journal}{Nature Electronics 2022 5:3}} \textbf{\bibinfo{volume}{5}}, \bibinfo{pages}{184--190} (\bibinfo{year}{2022}).
\newblock \urlprefix\url{https://www.nature.com/articles/s41928-022-00727-9}.

\bibitem{maurand_cmos_2016-1}
\bibinfo{author}{Maurand, R.} \emph{et~al.}
\newblock \bibinfo{title}{A {CMOS} silicon spin qubit}.
\newblock \emph{\bibinfo{journal}{Nature Communications}} \textbf{\bibinfo{volume}{7}}, \bibinfo{pages}{13575} (\bibinfo{year}{2016}).
\newblock \urlprefix\url{https://www.nature.com/articles/ncomms13575}.

\bibitem{camenzind_hole_2022}
\bibinfo{author}{Camenzind, L.~C.} \emph{et~al.}
\newblock \bibinfo{title}{A hole spin qubit in a fin field-effect transistor above 4 kelvin}.
\newblock \emph{\bibinfo{journal}{Nature Electronics}} \textbf{\bibinfo{volume}{5}}, \bibinfo{pages}{178--183} (\bibinfo{year}{2022}).
\newblock \urlprefix\url{https://www.nature.com/articles/s41928-022-00722-0}.

\bibitem{saraiva_materials_2022}
\bibinfo{author}{Saraiva, A.} \emph{et~al.}
\newblock \bibinfo{title}{Materials for {Silicon} {Quantum} {Dots} and their {Impact} on {Electron} {Spin} {Qubits}}.
\newblock \emph{\bibinfo{journal}{Advanced Functional Materials}} \textbf{\bibinfo{volume}{32}}, \bibinfo{pages}{2105488} (\bibinfo{year}{2022}).

\bibitem{elsayed_low_2024}
\bibinfo{author}{Elsayed, A.} \emph{et~al.}
\newblock \bibinfo{title}{Low charge noise quantum dots with industrial {CMOS} manufacturing}.
\newblock \emph{\bibinfo{journal}{npj Quantum Information}} \textbf{\bibinfo{volume}{10}}, \bibinfo{pages}{1--9} (\bibinfo{year}{2024}).
\newblock \urlprefix\url{https://www.nature.com/articles/s41534-024-00864-3}.

\bibitem{neyens_probing_2024}
\bibinfo{author}{Neyens, S.} \emph{et~al.}
\newblock \bibinfo{title}{Probing single electrons across 300-mm spin qubit wafers}.
\newblock \emph{\bibinfo{journal}{Nature}} \textbf{\bibinfo{volume}{629}}, \bibinfo{pages}{80--85} (\bibinfo{year}{2024}).
\newblock \urlprefix\url{https://www.nature.com/articles/s41586-024-07275-6}.

\bibitem{dumoulin_stuyck_demonstration_2024}
\bibinfo{author}{Dumoulin~Stuyck, N.} \emph{et~al.}
\newblock \bibinfo{title}{Demonstration of 99.9\% single qubit control fidelity of a silicon quantum dot spin qubit made in a 300 mm foundry process}.
\newblock In \emph{\bibinfo{booktitle}{2024 {IEEE} {Silicon} {Nanoelectronics} {Workshop} ({SNW})}}, \bibinfo{pages}{11--12} (\bibinfo{year}{2024}).
\newblock \urlprefix\url{https://ieeexplore.ieee.org/document/10639218/?arnumber=10639218}.

\bibitem{jock_silicon_2018}
\bibinfo{author}{Jock, R.~M.} \emph{et~al.}
\newblock \bibinfo{title}{A silicon metal-oxide-semiconductor electron spin-orbit qubit}.
\newblock \emph{\bibinfo{journal}{Nature Communications}} \textbf{\bibinfo{volume}{9}}, \bibinfo{pages}{1--20} (\bibinfo{year}{2018}).

\bibitem{tanttu_controlling_2019}
\bibinfo{author}{Tanttu, T.} \emph{et~al.}
\newblock \bibinfo{title}{Controlling {Spin}-{Orbit} {Interactions} in {Silicon} {Quantum} {Dots} {Using} {Magnetic} {Field} {Direction}}.
\newblock \emph{\bibinfo{journal}{Physical Review X}} \textbf{\bibinfo{volume}{9}}, \bibinfo{pages}{21028} (\bibinfo{year}{2019}).
\newblock \urlprefix\url{https://doi.org/10.1103/PhysRevX.9.021028}.

\bibitem{burkard_semiconductor_2023}
\bibinfo{author}{Burkard, G.}, \bibinfo{author}{Ladd, T.~D.}, \bibinfo{author}{Pan, A.}, \bibinfo{author}{Nichol, J.~M.} \& \bibinfo{author}{Petta, J.~R.}
\newblock \bibinfo{title}{Semiconductor spin qubits}.
\newblock \emph{\bibinfo{journal}{Reviews of Modern Physics}} \textbf{\bibinfo{volume}{95}}, \bibinfo{pages}{025003} (\bibinfo{year}{2023}).
\newblock \urlprefix\url{https://link.aps.org/doi/10.1103/RevModPhys.95.025003}.

\bibitem{cifuentes_impact_2024}
\bibinfo{author}{Cifuentes, J.~D.} \emph{et~al.}
\newblock \bibinfo{title}{Impact of electrostatic crosstalk on spin qubits in dense {CMOS} quantum dot arrays}.
\newblock \emph{\bibinfo{journal}{Physical Review B}} \textbf{\bibinfo{volume}{110}}, \bibinfo{pages}{125414} (\bibinfo{year}{2024}).
\newblock \urlprefix\url{https://link.aps.org/doi/10.1103/PhysRevB.110.125414}.

\bibitem{lawrie_quantum_2020}
\bibinfo{author}{Lawrie, W. I.~L.} \emph{et~al.}
\newblock \bibinfo{title}{Quantum dot arrays in silicon and germanium}.
\newblock \emph{\bibinfo{journal}{Applied Physics Letters}} \textbf{\bibinfo{volume}{116}}, \bibinfo{pages}{080501} (\bibinfo{year}{2020}).
\newblock \urlprefix\url{https://pubs.aip.org/apl/article/116/8/080501/38569/Quantum-dot-arrays-in-silicon-and-germanium}.

\bibitem{scappucci_germanium_2021}
\bibinfo{author}{Scappucci, G.} \emph{et~al.}
\newblock \bibinfo{title}{The germanium quantum information route}.
\newblock \emph{\bibinfo{journal}{Nature Reviews Materials}} \textbf{\bibinfo{volume}{6}}, \bibinfo{pages}{926--943} (\bibinfo{year}{2021}).
\newblock \urlprefix\url{https://www.nature.com/articles/s41578-020-00262-z}.

\bibitem{yang_spin-valley_2013}
\bibinfo{author}{Yang, C.~H.} \emph{et~al.}
\newblock \bibinfo{title}{Spin-valley lifetimes in a silicon quantum dot with tunable valley splitting}.
\newblock \emph{\bibinfo{journal}{Nature Communications}} \textbf{\bibinfo{volume}{4}} (\bibinfo{year}{2013}).

\bibitem{Cifuentes2024-du}
\bibinfo{author}{Cifuentes, J.~D.} \emph{et~al.}
\newblock \bibinfo{title}{Bounds to electron spin qubit variability for scalable {CMOS} architectures}.
\newblock \emph{\bibinfo{journal}{Nature Communications}} \textbf{\bibinfo{volume}{15}}, \bibinfo{pages}{4299} (\bibinfo{year}{2024}).

\bibitem{gonzalez-zalba_gate-sensing_2016}
\bibinfo{author}{Gonzalez-Zalba, M.~F.} \emph{et~al.}
\newblock \bibinfo{title}{Gate-{Sensing} {Coherent} {Charge} {Oscillations} in a {Silicon} {Field}-{Effect} {Transistor}}.
\newblock \emph{\bibinfo{journal}{Nano Letters}} \textbf{\bibinfo{volume}{16}}, \bibinfo{pages}{1614--1619} (\bibinfo{year}{2016}).

\bibitem{li_flexible_2020}
\bibinfo{author}{Li, R.} \emph{et~al.}
\newblock \bibinfo{title}{A flexible 300 mm integrated {Si} {MOS} platform for electron- and hole-spin qubits exploration}.
\newblock In \emph{\bibinfo{booktitle}{2020 {IEEE} {International} {Electron} {Devices} {Meeting} ({IEDM})}}, \bibinfo{pages}{38.3.1--38.3.4} (\bibinfo{year}{2020}).

\bibitem{fowler_surface_2012}
\bibinfo{author}{Fowler, A.~G.}, \bibinfo{author}{Mariantoni, M.}, \bibinfo{author}{Martinis, J.~M.} \& \bibinfo{author}{Cleland, A.~N.}
\newblock \bibinfo{title}{Surface codes: {Towards} practical large-scale quantum computation}.
\newblock \emph{\bibinfo{journal}{Physical Review A - Atomic, Molecular, and Optical Physics}} \textbf{\bibinfo{volume}{86}} (\bibinfo{year}{2012}).

\bibitem{shaw2022quantum}
\bibinfo{author}{Shaw, A.}, \bibinfo{author}{Gavriel, J.}, \bibinfo{author}{Herr, D.} \& \bibinfo{author}{Montanaro, A.}
\newblock \bibinfo{title}{Quantum computation on a 19-qubit wide 2d nearest neighbour qubit array}  (\bibinfo{year}{2022}).
\newblock \urlprefix\url{https://arxiv.org/abs/2212.01550}.
\newblock \eprint{2212.01550}.

\bibitem{blume-kohout_demonstration_2017}
\bibinfo{author}{Blume-Kohout, R.} \emph{et~al.}
\newblock \bibinfo{title}{Demonstration of qubit operations below a rigorous fault tolerance threshold with gate set tomography}.
\newblock \emph{\bibinfo{journal}{Nature Communications}} \textbf{\bibinfo{volume}{8}}, \bibinfo{pages}{14485} (\bibinfo{year}{2017}).
\newblock \urlprefix\url{https://www.nature.com/articles/ncomms14485}.

\bibitem{nielsen_probing_2020}
\bibinfo{author}{Nielsen, E.} \emph{et~al.}
\newblock \bibinfo{title}{Probing quantum processor performance with {pyGSTi}}.
\newblock \emph{\bibinfo{journal}{Quantum Science and Technology}} \textbf{\bibinfo{volume}{5}}, \bibinfo{pages}{044002} (\bibinfo{year}{2020}).
\newblock \urlprefix\url{https://dx.doi.org/10.1088/2058-9565/ab8aa4}.

\bibitem{dumoulin_stuyck_integrated_2020}
\bibinfo{author}{Dumoulin~Stuyck, N.} \emph{et~al.}
\newblock \bibinfo{title}{An {Integrated} {Silicon} {MOS} {Single}-{Electron} {Transistor} {Charge} {Sensor} for {Spin}-{Based} {Quantum} {Information} {Processing}}.
\newblock \emph{\bibinfo{journal}{IEEE Electron Device Letters}} \textbf{\bibinfo{volume}{41}}, \bibinfo{pages}{1253--1256} (\bibinfo{year}{2020}).

\bibitem{dumoulin_stuyck_uniform_2021}
\bibinfo{author}{Dumoulin~Stuyck, N.~I.} \emph{et~al.}
\newblock \bibinfo{title}{Uniform {Spin} {Qubit} {Devices} with {Tunable} {Coupling} in an {All}-{Silicon} 300 mm {Integrated} {Process}}.
\newblock \emph{\bibinfo{journal}{IEEE Symposium on VLSI Circuits, Digest of Technical Papers}} \textbf{\bibinfo{volume}{2021-June}} (\bibinfo{year}{2021}).

\bibitem{vahapoglu_single-electron_2021}
\bibinfo{author}{Vahapoglu, E.} \emph{et~al.}
\newblock \bibinfo{title}{Single-electron spin resonance in a nanoelectronic device using a global field}.
\newblock \emph{\bibinfo{journal}{Science Advances}} \textbf{\bibinfo{volume}{7}}, \bibinfo{pages}{9158--9171} (\bibinfo{year}{2021}).
\newblock \urlprefix\url{https://www.science.org}.

\bibitem{vahapoglu_coherent_2022}
\bibinfo{author}{Vahapoglu, E.} \emph{et~al.}
\newblock \bibinfo{title}{Coherent control of electron spin qubits in silicon using a global field}.
\newblock \emph{\bibinfo{journal}{npj Quantum Information}} \textbf{\bibinfo{volume}{8}}, \bibinfo{pages}{1--6} (\bibinfo{year}{2022}).
\newblock \urlprefix\url{https://www.nature.com/articles/s41534-022-00645-w}.

\bibitem{Chittock-Wood2024-hp}
\bibinfo{author}{Chittock-Wood, J.~F.} \emph{et~al.}
\newblock \bibinfo{title}{Exchange control in a mos double quantum dot made using a 300 mm wafer process} (\bibinfo{year}{2024}).
\newblock \urlprefix\url{https://arxiv.org/abs/2408.01241}.
\newblock \eprint{arXiv:2408.01241}.

\bibitem{seedhouse_pauli_2021}
\bibinfo{author}{Seedhouse, A.~E.} \emph{et~al.}
\newblock \bibinfo{title}{Pauli {Blockade} in {Silicon} {Quantum} {Dots} with {Spin}-{Orbit} {Control}}.
\newblock \emph{\bibinfo{journal}{PRX Quantum}} \textbf{\bibinfo{volume}{2}}, \bibinfo{pages}{010303} (\bibinfo{year}{2021}).
\newblock \urlprefix\url{https://link.aps.org/doi/10.1103/PRXQuantum.2.010303}.

\bibitem{vandersypen_nmr_2005}
\bibinfo{author}{Vandersypen, L. M.~K.} \& \bibinfo{author}{Chuang, I.~L.}
\newblock \bibinfo{title}{{NMR} techniques for quantum control and computation}.
\newblock \emph{\bibinfo{journal}{Reviews of Modern Physics}} \textbf{\bibinfo{volume}{76}}, \bibinfo{pages}{1037--1069} (\bibinfo{year}{2005}).
\newblock \urlprefix\url{https://link.aps.org/doi/10.1103/RevModPhys.76.1037}.

\bibitem{dumoulin_stuyck_silicon_2024}
\bibinfo{author}{Dumoulin~Stuyck, N.} \emph{et~al.}
\newblock \bibinfo{title}{Silicon spin qubit noise characterization using real-time feedback protocols and wavelet analysis}.
\newblock \emph{\bibinfo{journal}{Applied Physics Letters}} \textbf{\bibinfo{volume}{124}}, \bibinfo{pages}{114003} (\bibinfo{year}{2024}).
\newblock \urlprefix\url{https://doi.org/10.1063/5.0179958}.

\bibitem{Reed2016}
\bibinfo{author}{Reed, M.~D.} \emph{et~al.}
\newblock \bibinfo{title}{Reduced {Sensitivity} to {Charge} {Noise} in {Semiconductor} {Spin} {Qubits} via {Symmetric} {Operation}}.
\newblock \emph{\bibinfo{journal}{Physical Review Letters}} \textbf{\bibinfo{volume}{116}}, \bibinfo{pages}{1--6} (\bibinfo{year}{2016}).

\bibitem{steinacker_violating_2024}
\bibinfo{author}{Steinacker, P.} \emph{et~al.}
\newblock \bibinfo{title}{Violating {Bell}'s inequality in gate-defined quantum dots} (\bibinfo{year}{2024}).
\newblock \urlprefix\url{http://arxiv.org/abs/2407.15778}.

\bibitem{Nielsen2021gatesettomography}
\bibinfo{author}{Nielsen, E.} \emph{et~al.}
\newblock \bibinfo{title}{Gate {S}et {T}omography}.
\newblock \emph{\bibinfo{journal}{{Quantum}}} \textbf{\bibinfo{volume}{5}}, \bibinfo{pages}{557} (\bibinfo{year}{2021}).
\newblock \urlprefix\url{https://doi.org/10.22331/q-2021-10-05-557}.

\bibitem{Muhonen2014}
\bibinfo{author}{Muhonen, J.~T.} \emph{et~al.}
\newblock \bibinfo{title}{Storing quantum information for 30 seconds in a nanoelectronic device}.
\newblock \emph{\bibinfo{journal}{Nature Nanotechnology}} \textbf{\bibinfo{volume}{9}}, \bibinfo{pages}{986--991} (\bibinfo{year}{2014}).
\newblock \urlprefix\url{www.nature.com/naturenanotechnology}.

\bibitem{hensen_silicon_2020}
\bibinfo{author}{Hensen, B.} \emph{et~al.}
\newblock \bibinfo{title}{A silicon quantum-dot-coupled nuclear spin qubit}.
\newblock \emph{\bibinfo{journal}{Nature Nanotechnology}} \textbf{\bibinfo{volume}{15}}, \bibinfo{pages}{13--17} (\bibinfo{year}{2020}).
\newblock \urlprefix\url{https://www.nature.com/articles/s41565-019-0587-7}.

\bibitem{zhao_single-spin_2020}
\bibinfo{author}{Zhao, R.} \emph{et~al.}
\newblock \bibinfo{title}{Single-spin qubits in isotopically enriched silicon at low magnetic field}.
\newblock \emph{\bibinfo{journal}{Nature Communications}} \textbf{\bibinfo{volume}{10}}, \bibinfo{pages}{5500} (\bibinfo{year}{2019}).
\newblock \urlprefix\url{https://doi.org/10.1038/s41467-019-13416-7}.

\bibitem{Struck2020}
\bibinfo{author}{Struck, T.} \emph{et~al.}
\newblock \bibinfo{title}{Low-frequency spin qubit energy splitting noise in highly purified {28Si}/{SiGe}}.
\newblock \emph{\bibinfo{journal}{npj Quantum Information}} \textbf{\bibinfo{volume}{6}} (\bibinfo{year}{2020}).
\newblock \urlprefix\url{https://doi.org/10.1038/s41534-020-0276-2}.

\bibitem{Yoneda2018}
\bibinfo{author}{Yoneda, J.} \emph{et~al.}
\newblock \bibinfo{title}{A quantum-dot spin qubit with coherence limited by charge noise and fidelity higher than 99.9\%}.
\newblock \emph{\bibinfo{journal}{Nature Nanotechnology}} \textbf{\bibinfo{volume}{13}}, \bibinfo{pages}{102--106} (\bibinfo{year}{2018}).
\newblock \urlprefix\url{http://dx.doi.org/10.1038/s41565-017-0014-x}.

\bibitem{tyryshkin_electron_2012}
\bibinfo{author}{Tyryshkin, A.~M.} \emph{et~al.}
\newblock \bibinfo{title}{Electron spin coherence exceeding seconds in high-purity silicon}.
\newblock \emph{\bibinfo{journal}{Nature Materials}} \textbf{\bibinfo{volume}{11}}, \bibinfo{pages}{143--147} (\bibinfo{year}{2012}).
\newblock \urlprefix\url{http://dx.doi.org/10.1038/nmat3182}.

\bibitem{camenzind_high_2021}
\bibinfo{author}{Camenzind, T.~N.} \emph{et~al.}
\newblock \bibinfo{title}{High mobility {SiMOSFETs} fabricated in a full 300 mm {CMOS} process}.
\newblock \emph{\bibinfo{journal}{Materials for Quantum Technology}} \textbf{\bibinfo{volume}{1}}, \bibinfo{pages}{041001} (\bibinfo{year}{2021}).
\newblock \urlprefix\url{https://dx.doi.org/10.1088/2633-4356/ac40f4}.

\bibitem{bartee_spin_2024}
\bibinfo{author}{Bartee, S.~K.} \emph{et~al.}
\newblock \bibinfo{title}{Spin {Qubits} with {Scalable} milli-kelvin {CMOS} {Control}} (\bibinfo{year}{2024}).
\newblock \urlprefix\url{https://arxiv.org/abs/2407.15151v1}.

\bibitem{petit_spin_2018}
\bibinfo{author}{Petit, L.} \emph{et~al.}
\newblock \bibinfo{title}{Spin {Lifetime} and {Charge} {Noise} in {Hot} {Silicon} {Quantum} {Dot} {Qubits}}.
\newblock \emph{\bibinfo{journal}{Physical Review Letters}} \textbf{\bibinfo{volume}{121}}, \bibinfo{pages}{1--8} (\bibinfo{year}{2018}).

\end{thebibliography}

\section*{Methods}
\setcounter{section}{0}

\subsection{Measurement setup}\label{methods:measurement_setup}
The device is measured in a Bluefors XLD400 dilution refrigerator and mounted on the cold finger. An external DC magnetic field is supplied by an American Magnetics AMI430 magnet points in the [110] direction of the Si lattice. DC voltages are supplied with a Q-Devil QDAC-II, through DC lines with a bandwidth from 0 to $\SI{20}{\hertz}$. Dynamic voltage pulses are generated with a Quantum Machines (QM) Operator-X+ (OPX+) and combined with DC biases using custom linear bias combiners at room temperature. The OPX+ has a sampling time of $\SI{4}{\nano\second}$. The dynamic pulse lines in the fridge have a bandwidth of 0 to $\SI{50}{\mega\hertz}$, which translates into a minimum rise time of $\SI{20}{\nano\second}$. Microwave pulses are generated by a Keysight PSG8267D Vector Signal Generator, with in-phase and quadrature (I/Q) and pulse modulation waveforms generated by the QM OPX+.

The charge sensor comprises a single-island DC-SET. The SET current integrated for $t_\mathrm{int} = \SI{100}{\micro \second}$ is amplified using a room-temperature I–V converter (Basel SP983c) and sampled by a QM OPX+. 




\section{Data Availability}
The datasets generated and/or analyzed during this study are available from the corresponding authors on reasonable request.

\section*{Acknowledgements}
\noindent We thank C\'edric Boh\'emier for his assistance in cryogenic screening. We acknowledge support from the Australian Research Council (FL190100167). P. S. and A. N. acknowledge support from the Sydney Quantum Academy. P. S. acknowledges support from the Baxter Charitable Foundation.

\subsection{Author Contributions}
W. H. L., K. W. C., F. E. H., C. C. E. and N. D. S. designed the device.
P. S. conducted the experiments with N. D. S.'s supervision and input from T. T., A. S., A. L., C. H. Y., M. F., A. N., J. D. C., C. C. E. and A. S. D..
S. S. assisted with the experimental setup.
N. D. S. wrote the GST experiment implementation. 
M. F. helped with the GST analysis. 
The imec team developed the $\SI{300}{\milli\meter}$ spin qubit process, fabricated the device, and performed an initial electrical device screening at wafer-scale.
M. C. performed initial cryogenic device screening characterisation with W. H. L.'s supervision.
N. D. S., P. S. and A. S. wrote the manuscript, with input from all authors.


\section{Competing Interests}
A. S. D. is CEO and a director of Diraq Pty Ltd. N. D. S, T. T., W. H. L., K. W. C., C. C. E., F. E. H., A. L., C. H. Y., A. S. and A. S. D. declare equity interest in Diraq. Other authors declare no competing interest.

\setcounter{figure}{0}
\setcounter{table}{0}
\captionsetup[figure]{name={\bf{Extended Data Fig.}},labelsep=line,justification=raggedright,font=small,singlelinecheck=false}

\begin{figure*}[ht]
    \includesvg[width = \textwidth]{sfig1.svg}
    \caption{\textbf{Single qubit metrics} 
    \textbf{a,b}, Spin relaxation of qubit 1 (\textbf{a}) and 2 ({\textbf{b}}) spin up to the ground state $\ket{\downarrow\downarrow}$. Curves are fitted to an exponential decay resulting in a relaxation time of $T_\mathrm{1,Q1} = \SI{2.4\pm 0.2}{\second}$ and $T_\mathrm{1,Q2} = \SI{6.3\pm 0.6}{\second}$, respectively.
    \textbf{c,d}, Ramsey experiment for qubit 1 (\textbf{c}) and 2 ({\textbf{d}}) fitted to a sinusoidal exponential decay resulting in a coherence time of $T_\mathrm{2,Q1}^{*} = \SI{30.4\pm 0.8}{\micro \second}$ and $T_\mathrm{2,Q2}^{*} = \SI{29.1\pm 0.6}{\micro \second}$, respectively. The oscillation is induced by a phase shift corresponding to a \SI{1}{\mega \hertz} detuning. Every data point is averaged for 1000 repeats, each integrating the readout signal for $t_\mathrm{int} = \SI{100}{\micro \second}$. We use real-time Larmor frequency tracking between repeats using the protocol described in Ref.~\cite{dumoulin_stuyck_silicon_2024}.
    \textbf{e,f}, Hahn echo experiment for qubit 1 (\textbf{e}) and 2 ({\textbf{f}}) fitted to an exponential decay resulting in a coherence time of $T_\mathrm{2,Q1}^{\mathrm{Hahn}} = \SI{445\pm 6}{\micro \second}$ and $T_\mathrm{2,Q2}^{\mathrm{Hahn}} = \SI{803\pm 6}{\micro \second}$, respectively. In this experiment, we measure all six single qubit projections to fit the state purity $\gamma_\mathrm{state}$.
    Error bars represent the \SI{95}{\percent} confidence level.
}
    \label{fig:sup_fig1}
\end{figure*}

\begin{figure*}[ht]
    \includesvg[width = \textwidth]{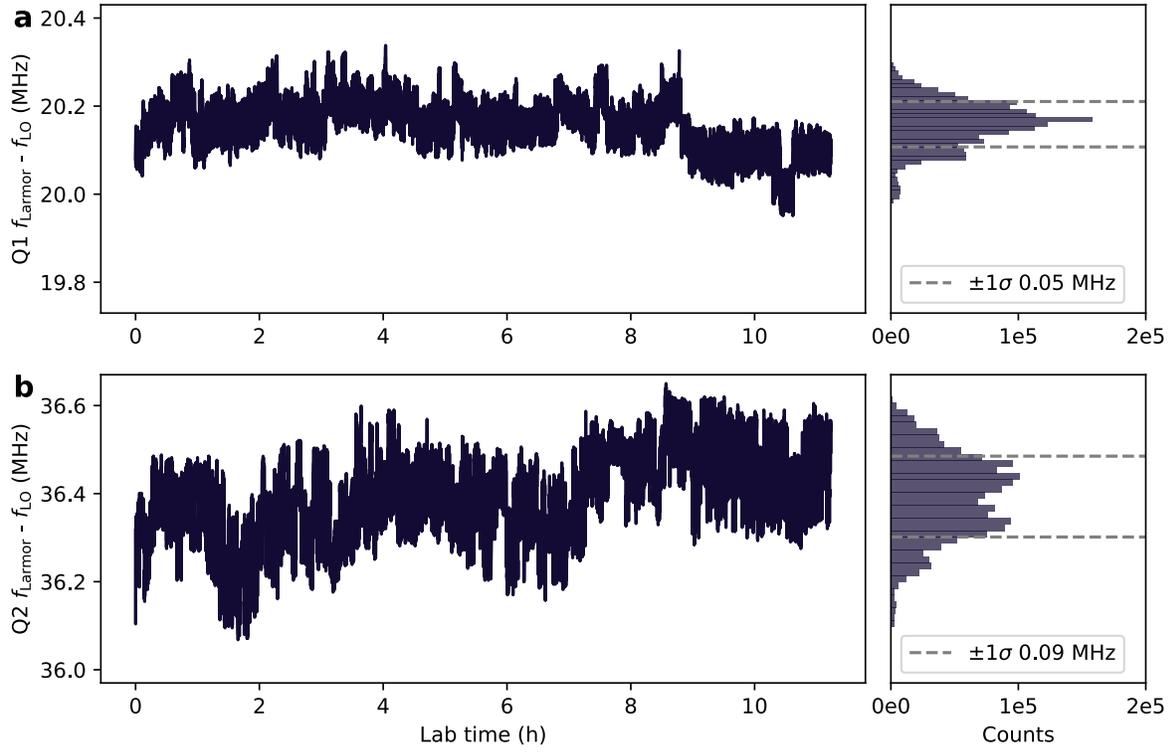}
    \caption{\textbf{Single qubit Larmor frequency tracking.} 
    \textbf{a,b} Single qubit Larmor frequency deviation as a function of lab time during the GST experiment discussed in the main text for qubit 1 and 2, respectively. Figures are plotted with the same y-axis range. Qubit 2 has a frequency standard deviation of $\SI{0.09}{\mega\hertz}$ compared to $\SI{0.05}{\mega\hertz}$ for qubit 1. Qubit 2 also shows signs of two discrete frequency levels in the histogram at $\approx \SI{36.4}{\mega\hertz}$ and $\SI{36.3}{\mega\hertz}$. The local oscillator frequency is set to $f_{\mathrm{LO}} = \SI{18.610}{\giga \hertz}$. Frequencies are tracked using the protocol described in Ref.~\cite{dumoulin_stuyck_silicon_2024}.
}
    \label{fig:sup_fig2}
\end{figure*}
\begin{figure*}[ht]
    \includesvg[width = 0.6\textwidth]{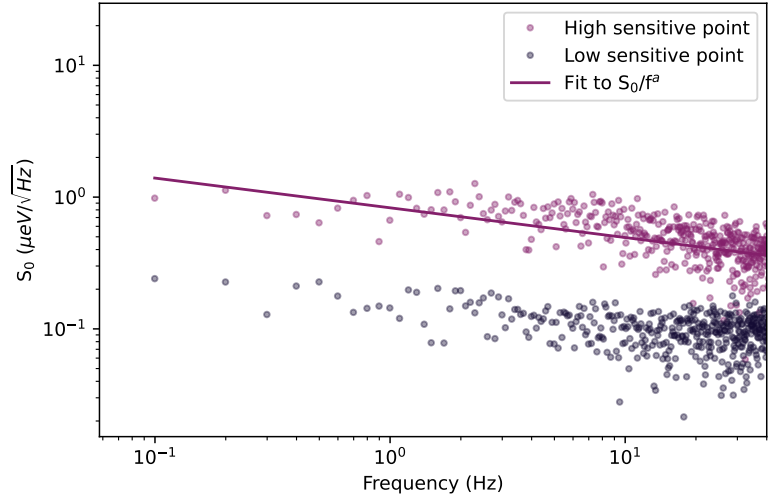}
    \caption{\textbf{Single electron transistor charge noise spectrum.} Noise spectrum at a high and a low-sensitive point of a Coulomb oscillation. Voltage noise spectrum is first calculated using a Fast Fourier transform on extracted SET current measurements divided by the slope of the Coulomb peak, and finally converted into a charge noise spectrum in $\SI{}{\micro\electronvolt}$ using the lever arm~\cite{petit_spin_2018}. Charge noise at $\SI{1}{\hertz} = 0.83(1)\SI{}{\micro\electronvolt\per \sqrt{\hertz}}$. Solid line is a power law fit with exponent $a = 0.226(2)$.
}
    \label{fig:sup_fig3}
\end{figure*}

\begin{figure*}[ht]
    \includesvg[width = 1.0\textwidth]{sfig4.svg}
    \caption{\textbf{State preparation and measurement matrix.} 
    GST matrices of $\ket{\downarrow \downarrow}$ state initialization and even and odd parity measurement result in a fidelity of $F_\mathrm{init} = \SI{99.996\pm 0.049}{\percent}$ and $F_\mathrm{meas} = \SI{99.959\pm 0.059}{\percent}$, respectively. The combined state preparation and measurement fidelity is calculated to $F_\mathrm{SPAM} = \SI{99.95\pm 0.08}{\percent}$. Error bars represent the \SI{95}{\percent} confidence level.
}
    \label{fig:sup_fig4}
\end{figure*}

\end{document}